# Contrasting results of surface metrology techniques for three-dimensional human fingerprints

Brian Lee Beatty[1]

Shani Kahan[1]

Burcak Bas[1]

Bettina Zou[1]

Nicole Werpachowski[1]

[1]NYIT College of Osteopathic Medicine, Northern Blvd, Old Westbury, New York 11568, USA

Abstract

Fingerprints, otherwise known as dermatoglyphs, are most commonly thought of in the context of identification, but have myriad other roles in human biology. They are formed by the restricted ability of ridges and furrows of the epidermis to flatten. The patterns these ridges and furrows make can be represented as 2D fingerprints, but also as 3D structures with cross-sectional shapes that may add new levels of detail to identification, forensic, and behavioral uses/studies. Surface metrology techniques better allow for the quantification of these features, though it is unclear what tool and what scale is most appropriate. A Sensofar S Neox white light reflectance confocal microscope and a Gelsight Mobile 2 were used to independently measure the surface roughness of the fingerprints of four individuals from preserved cadaveric remains. Scale-sensitive fractal analyses (SSFA) were performed on the data from the S Neox (a small area), Gelsight (a larger area), and the same Gelsight datasets cropped down to the size of the S

Neox scan size. Though fewer SSFA parameters identified differences between individuals from the smaller, extracted Gelsight area, all three forms of measurement found significant differences between some individuals from the study. No significant differences were found that differ between fingers themselves. Though only an initial step, these data suggest that a variety of surface metrology techniques may be useful in differentiating individuals.

**Introduction**

Fingerprints (aka dermatoglyphs) are typically seen as two-dimensional pictures of parallel lines curving around the surface of a fingertip. This is an obvious illusion, as it is well-known that these lines are a simplified representation of the ridges separated by furrows that make the three-dimensional surface of the glabrous skin of fingertips. Despite this, fingerprints as a means of identification remain a 2D representation, originally developed as a means of identification most notably by Jan Evangelista Purkinje in the 1800s (Grzybowski & Pietrzak, 2014) and remains an important means of identification to the present day. The two-dimensional representation of dermatoglyphs captures the curvature of the ridges and their connectivity, and the patterns that these create. Yet the two-dimensional representation fails to capture the three-dimensional shapes of the ridges and furrows themselves, including the curvature and depths of their profiles.

Three-dimensional scans of fingerprints are not a new concept though older forms focus more on the challenges of converting 3D scans to 2D fingerprints, measuring the presence of ridges without capturing the cross-sectional shapes of ridges (Chen et al., 2006; Zhao et al., 2011; Pang & Song, 2013; Kumar et al., 2015; Liu et al., 2015). Surprisingly, even newer studies of 3D fingerprinting are focused on conversion of 3D to 2D (Cui et al., 2023). But the ability to capture three-dimensionality of fingerprints offers much more information than simple conversion to two dimensions, including the heights and depths of ridges and furrows, which might break down with damage to their underlying structures that keep them rigid, such as exposure to heat or manual labor. Plus, the cross-sectional profiles of the ridges and furrows, including the ways these cross sectional shapes curve and/or undulate, may reveal details that are unique to individuals or the result of specific behaviors and occupations (Batool & Izzat, 2023). But without investigating these details, how will we ever know?

Surface metrology, whether from non-contact methods that utilize reflected light like those used in recent studies of the human dorsal nail plate (Beatty et al., 2024), or newer methods that use contact with a deformable elastomer such as Gelsight (Stewart et al., 2024), can easily solve the problem of capturing these data on three-dimensional features. The analysis of such geometries is much easier to characterize since the development of scale-sensitive fractal analysis (SSFA) (Brown et al., 2018). Collectively, the tools and techniques for measurement and analyses of surface roughness have advanced greatly over the past decade, and it appears that treating fingerprints as the three-dimensional structures they are is long overdue.

Non-contact white light reflectance confocal microscopy (in this case, with a Sensofar S Neox optical profiler) requires light to reflect from a surface in an approximately uniform way. Prior studies measuring complex surfaces with this instrument have shown clear epoxy casts to be reliably good for such scans (Mihlbachler et al, 2022). The S Neox, as well as other tools of this ilk, have options for multiple microscope objectives that make it possible to measure smaller and smaller areas at finer and finer scales, down to sub-micron resolutions. Such scans are susceptible to problems from uneven reflectivity, and typically take minutes or hours to scan, depending on the size and resolution desired.

The Gelsight Mobile 2 is a tool that utilizes the formation of a 3D point cloud from photogrammetry principles applied to the deformation of an elastomer as it is pressed against a surface. Because the Gelsight Mobile 2 requires contact and small pressures, softer tissues can be deformed themselves, by this pressure, altering the natural texture. The elastomer and detector themselves are not yet alterable in size or resolution, making this tool less flexible in scale of measurement as well. But the Gelsight Mobile 2 takes a second to collect data, providing a tradeoff that might make it more amenable to studies of living patients.

It is challenging to know what tools and what scales are best for analyzing fingerprints. Beyond this initial study there is surely more to do, but at the outset it would be best to recognize tools and measures that are significantly different between individuals, and randomly variable between fingers of the same hand. This study investigates this among cadaveric individuals for two reasons: 1) simply put, cadaveric individual identities and privacy have simpler requirements for a pilot study, and 2) cadaveric fingerprints might better represent individuals in more poorly preserved circumstances, common in forensics, where identification with fingerprints might be useful. Using the S Neox and Gelsight Mobile 2, this study aimed at recognizing which tools and which scales are best at finding significant differences between individuals. It is likely that when the data from the S Neox is measured at the same scale as the Gelsight Mobile 2, these will both find measurable differences between individuals that largely overlap. The Gelsight Mobile 2 captures a larger area than the lowest magnification of the S Neox, making for a readily straightforward comparison available to examine the large area originally scanned by Gelsight, the smaller area scanned by the S Neox, and the Gelsight area cropped to the smaller size of the S Neox.

**Materials & Methods**

Sampling

Four individual whole body donors were chosen for the visibly clean and intact quality of their fingerprints. The identities of the individuals were not recorded, and each was given a simple number identifier so that they could be recognized within the study. Because these specimens do not meet the definition of a human subject according to the IRB, they do not require IRB exempt review. The study was HIPAA compliant and adhered to the tenets of the Declaration of Helsinki. The affiliation of the authors with License for Tissue Bank Operation (DOH 3908(04/2001)) has the consent for using tissue from donor and consent for scientific research and article publishing.

Each finger was cleaned with alcohol wipes and allowed to dry for 6 hours, then molded with President Jet regular body polyvinylsiloxane. The molds were then prepared for casting by building a retaining wall around each mold with polyvinylsiloxane putty (also by Coltene Whaledent President Jet). Casts were made with clear epoxy resin (Epokwik, Buehler) and allowed to cure for 24 hours.

Scanning

These clear epoxy resin casts were scanned with the Gelsight Mobile 2 and Sensofar S Neox in approximately the same locations. The S Neox scan utilized a blend of confocal microscopy and focus variation (termed, "confocal fusion"). The 5x objective produced a scan area that was 3.49mm x 2.63mm (Figure 1).

Gelsight Mobile 2 scans, which are initially at a much larger scale, were done aimed at the same location, with moderately applied pressure so that the scan captured as much data with minimal draping as possible (Figure 2). "Draping" is the phenomenon where the elastomer surface is stretched over deeper gaps in the surface being scanned, rather than filling them. Typically, pressure alleviates this problem, though the depth of the gap and the viscosity of the elastomer can result in unavoidable draping. Gelsight Mobile 2 scans produced a scan area that was approximately 1cm x 1.2cm. These original Gelsight Mobile 2 scans are hereafter referred to as Gelsight BIG, to differentiate them from Gelsight SMALL areas. Gelsight SMALL areas are derived from Gelsight BIG areas, simply extracted using SensoMap software (see below).

Analysis

Each scan, including S Neox 5x scans, Gelsight BIG, and Gelsight SMALL, were processed using SensoMap software (version 9.0, Sensofar Group, Barcelona, Spain). After S Neox 5x scans and Gelsight BIG scans were loaded, an area of 3.49mm x 2.63mm (identical to the S Neox scan size) was extracted from the Gelsight BIG scans and saved as the Gelsight SMALL scans. Processing starts with leveling and the form removal (polynomial 3) to minimize the effects of scans being tilted or of round surfaces for which we only want the texture of the ridges and furrows, not the tilt or roundness of the fingertip itself. Surfaces were retouched

(edited) for contamination and artifacts, such as bubbles on cast surfaces. Surfaces were leveled again to readjust in case the removal of artifacts or form had prevented proper leveling in the first round of leveling. After this basic data processing, SSFA parameters were derived for each of these surfaces using SensoMap software and saved as .xlsx spreadsheets.

*Statistics*

All surface metrology variables collected (Supplemental Materials) were analyzed using IBM SPSS Statistics (Version 28.0.1). Tests of normality found most parameters exhibited normal distributions, thus parametric tests (t-tests and ANOVA) were carried out. In prior studies using the same variables (Mihlbachler et al, 2022), no meaningful differences were found when analyzed with non-parametric tests. To identify sources of significant differences, post-hoc Tukey tests were used.

**Results**

*Sensofar S Neox*

A one-way ANOVA of the samples scanned with the S Neox found no significant differences in variables when SSFA parameters were grouped by finger number. Thus, there seems to be no reason to think that, for example, the dermatoglyphs from digit I are measurably different from digit II, or digit II from digit III, and so on.

On the other hand, when grouped by individual, significant differences were found in several of the SSFA parameters tested (Figure 3). The maximum relative distance (*Y max*) was significantly different ($p = .043$), with a Tukey's post hoc test indicating that the difference lay between individual #1 and #4. Similarly, the differences between individuals #1 and #4 were significantly different in the Smooth-rough crossover threshold (*SRC threshold*) (p = .043). Significant differences were also found for complexity (*Lsfc*) (p = .02) and fractal dimensions (*DIs*) (p = .02), but between individuals #1 and #4 as well as between individuals #2 and #4.

### Gelsight BIG area

A one-way ANOVA of the samples scanned for larger areas with Gelsight found no significant differences in variables when SSFA parameters were grouped by finger number.

On the other hand, when grouped by individual, significant differences were found in several of the SSFA parameters tested (Figure 4). The maximum relative distance (*Y max*) was significantly different (p = .013), with a Tukey's post hoc test indicating that the difference lay between individual #1 and #3 and #2 and #3.. Similarly, the differences between individuals #1 and #3 and #2 and #3 were significantly different in the Smooth-rough crossover threshold (*SRC threshold*) (p = .013), complexity (*Lsfc*) (p = .006) and fractal dimensions (*DIs*) (p = .006). *SRC* was significantly different (p = .027) between individuals #2 and #3.

### Gelsight SMALL area

Like the Sensofar S Neox and larger area scanned with Gelsight, a one-way ANOVA of the samples scanned for smaller areas with Gelsight found no significant differences in variables when SSFA parameters were grouped by finger number.

A one-way ANOVA of the samples scanned for smaller areas with Gelsight found some significant differences in variables when SSFA parameters were grouped by individual (Figure 5). The maximum relative distance (*Y max*) was significantly different (p = .047), with a Tukey's post hoc test indicating that the difference lay between individual #2 and #3. The differences between individuals #2 and #3 were significantly different in the Smooth-rough crossover threshold (*SRC threshold*) (p = .047).

**Discussion**

The Gelsight regions, both large and small, appear to exhibit more variables that differentiate individuals than does the S Neox data. Considering that the small Gelsight region is the same size as the S Neox region that was scanned, this does not seem to be a matter of the scale of the sampling region. Though it was not possible to control the regions scanned to make these two capture identical locations, these differences hint that differences in the nature of these two data collection techniques might explain this.

For example, the elastomer of the Gelsight Mobile 2 has a natural tendency for "draping", a phenomenon in which the elastomer is stretched to its maximum within a deep depression without completely filling the sharper angles of these depressions, leaving curved surfaces where the elastomer surface fails to meet the surface being scanned. The S Neox relies entirely on reflected light from the surface, so such draping is not a possibility. The confocal microscopy and focus variation methods the S Neox employs has its own limitations with narrow steep surfaces, but this is visibly not as severe an issue as it is with the Gelsight Mobile 2.

In contrast, the S Neox uses reflected light that can scatter if something is unusually reflective, and this scatter is often captured as random spikes in the surface. SensoMap 9 makes it easy to edit these out, yet these corrections might alter the surface in subtle ways that can add up. The Gelsight Mobile 2 elastomer is in contact with the surface, making such random spikes of noise from reflectivity not a possibility.

Thus, while the S Neox might fundamentally avoid draping and have greater potential resolution, the Gelsight Mobile 2 avoids problems of artifacts of unusual reflectivity and generally avoids data contaminated by spikes of noise. It is also noteworthy that the Gelsight Mobile 2 is a much faster means of collecting these data, and it is more readily used with living people. Much more than this preliminary evaluation is needed to determine which data form is best for studies of dermatoglyphs in the future, but there seems to be hope that it can be used to identify individuals, even from fixed cadaveric samples.

### *Forensics*

What might this mean for forensics? Genetics is a far more reliable and distinct way to identify individuals if one has enough tissue to measure a fingerprint. Yet there is one scenario where genetics might not assist the forensics specialist - twins. Genetically identical twins have identical genomes, yet their fingerprint patterns (and perhaps also the three-dimensional features of ridges and furrows of the dermatoglyph) differ as a result of the Turing reaction-diffusion system that produces the whorls and loops typical of fingerprints via development (Glover et al., 2023). If three-dimensional features of fingerprints could be further analyzed in twin studies, one could find further differences or similarities in their fingerprints that could aid or confound distinction between twins.

### *Congenital Disorders and Cutaneous Dermatoses*

Existing literature has elicited how dermatoglyphics can offer invaluable insights into an individual's health status, even indicating the presence of congenital disorders including Klinefelter's syndrome (Shiono et al., 1977), Down's syndrome, and Turner syndrome (Holt et al., 1964). This illuminates the potential for dermatoglyphics to serve as a non-invasive adjunct for the detection of various genetic disorders. For example, one study investigating palmar dermatoglyphics in Down's syndrome demonstrated unique patterns including ulnar loop patterns in fingertips and abnormal patterns in hypothenar and interdigital areas (Plato et al., 1973). Furthermore, another comparative study highlighted gender differences in dermatoglyphic patterns among individuals with schizophrenia, revealing higher total finger ridge count (and presumably smaller ridges and steeper borders between them) in males (Sivkov et al., 2009). These examples illustrate the broader potential of dermatoglyphics in medical diagnostics beyond biometric identification and the importance of continued research in this interdisciplinary field.

The influence of skin diseases on fingerprint recognition is a significant area of study, particularly in understanding how various conditions impact the histopathological structures critical for biometric identification. For instance, skin diseases including atopic dermatitis, verruca vulgaris, psoriasis, systemic lupus erythematosus, and epidermolysis bullosa have been shown to cause notable changes at the dermo-epidermal junction. This translates to the disruption of papillary lines, leading to the alteration or loss of identifiable fingerprint ridges, rendering fingerprints unsuitable for identification by pattern recognition (Drahansky et al., 2012). These findings underscore the critical intersection between dermatology and biometric security, highlighting the challenges and considerations in fingerprint recognition technology for individuals with skin diseases.

**Acknowledgements**

Kelsi Hurdle provided support and access to the Sensofar S Neox at the NYIT College of Osteopathic Medicine Visualization Center.

Table 1. Scale-sensitive fractal analysis parameters utilized in this study.

Figure captions

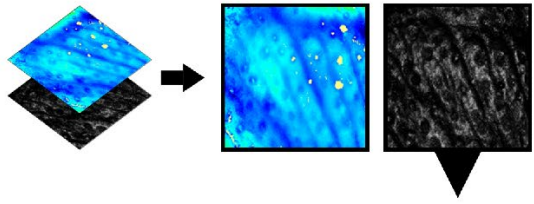

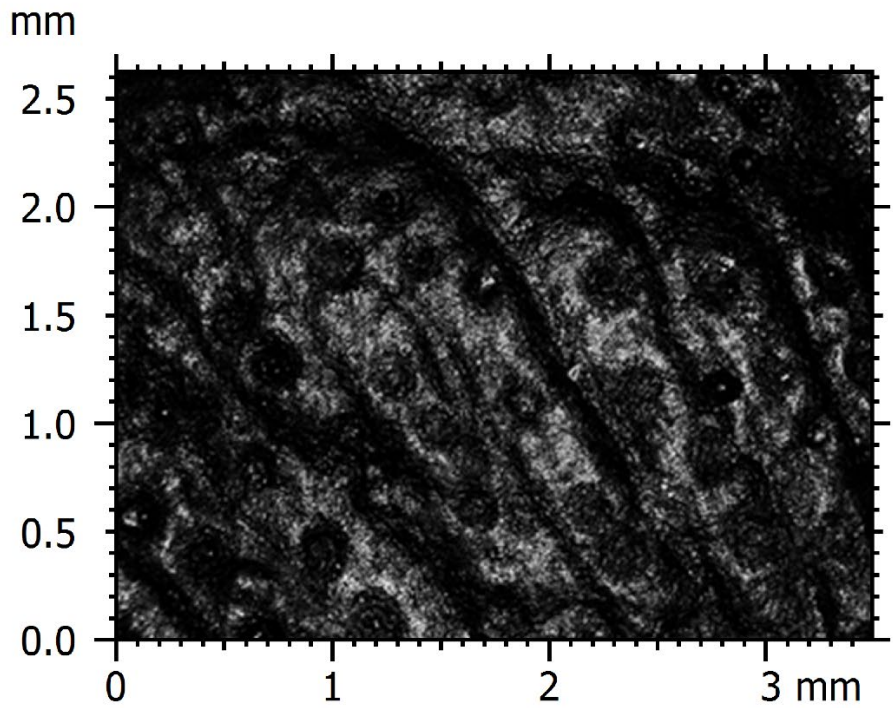

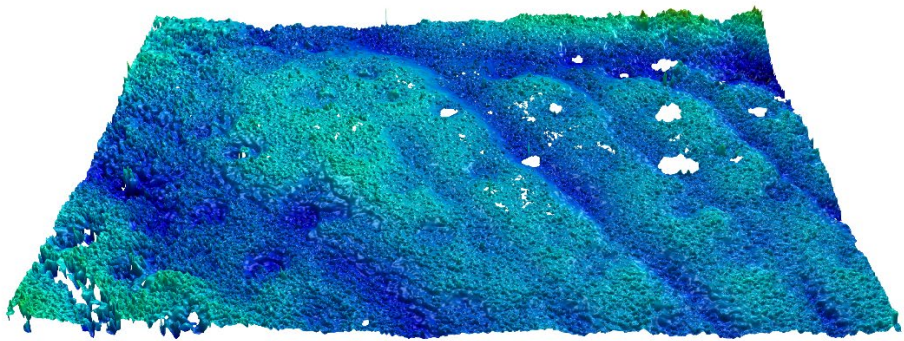

Figure 1. Sensofar S Neox 5x objective scan of a fingerprint, showing a grayscale image of the scan and a 3D representation of that scan. Note the scale in the top image and the recognizable ridges, sulci, and sweat pores in the bottom image.

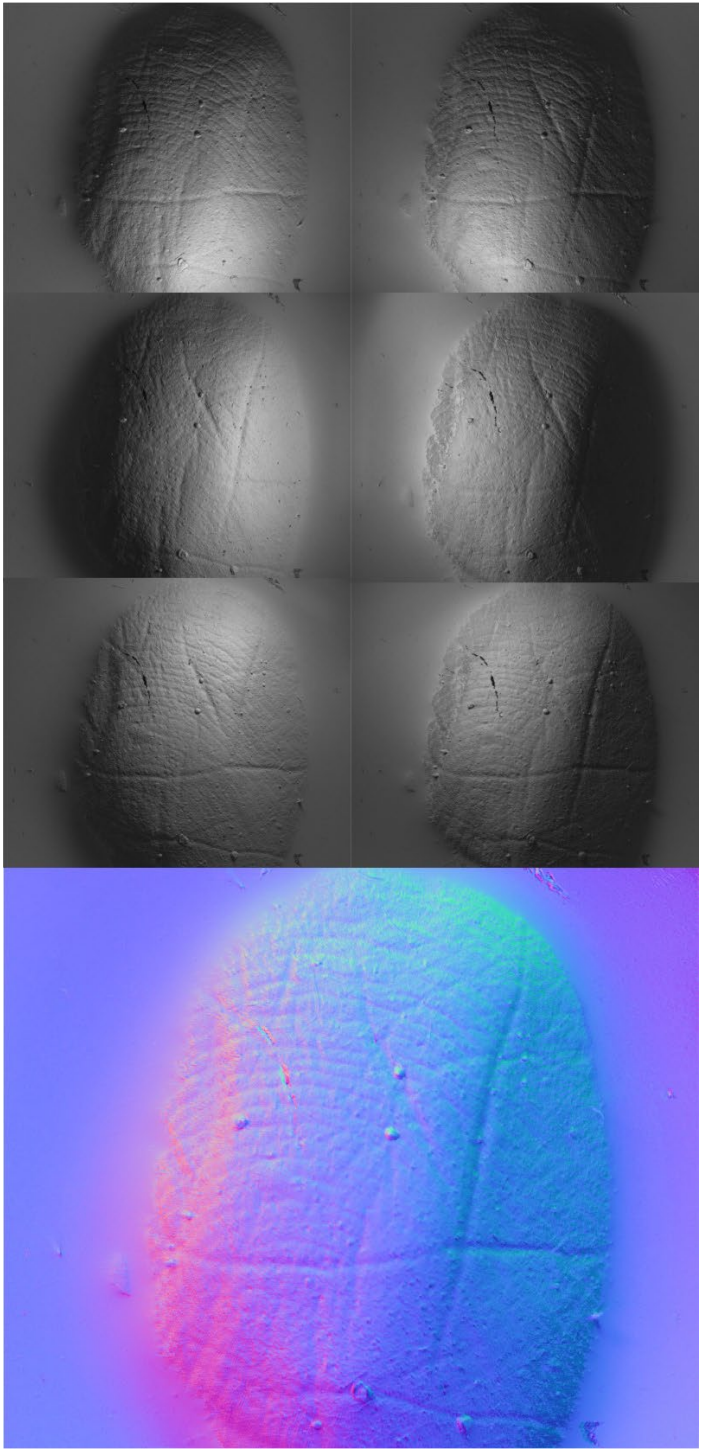

Figure 2. Gelsight scan of fingerprint, with six grayscale images from each of the six angles of lighting (top 3 rows); multicolor image representation of the 3D point cloud of data interpreted from the light and shadowed parts of the overlain six grayscale images (bottom).

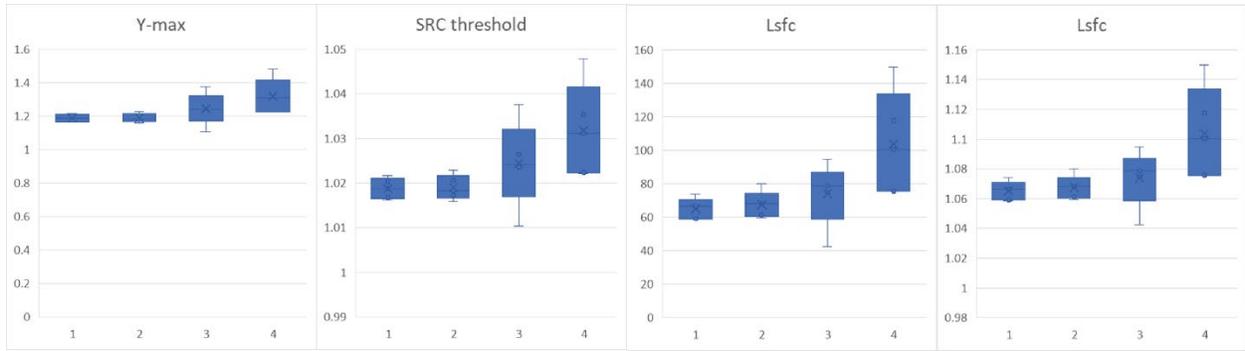

Figure 3. Significant SSFA parameters differing between individuals from Sensofar S Neox 5x data. Tukey's post hoc test identified these differences being between individuals #1 and #4 for maximum relative distance (*Y max*) (p = .043), Smooth-rough crossover threshold (*SRC threshold*) (p = .043), complexity (*Lsfc*) (p = .02) and fractal dimensions (*DIs*) (p = .02). Tukey's post hoc test also found differences in complexity (*Lsfc*) (p = .02) and fractal dimensions (*DIs*) (p = .02) between individuals #2 and #4.

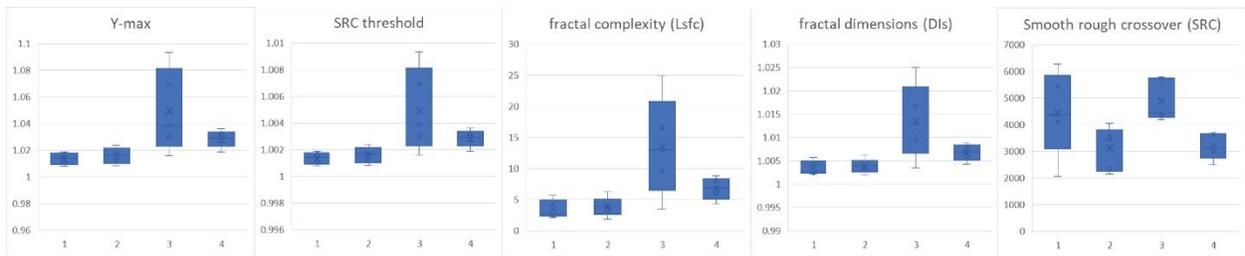

Figure 4. Significant SSFA parameters differing between individuals from Gelsight BIG scans. Tukey's post hoc test identified several parameters differing between individuals #1 and #3 and #2 and #3, including maximum relative distance (*Y max*) (p = .013), Smooth-rough crossover threshold (*SRC threshold*) (p = .013), complexity (*Lsfc*) (p = .006) and fractal dimensions (*DIs*) (p = .006). *SRC* was also significantly different (p = .027) between individuals #2 and #3.

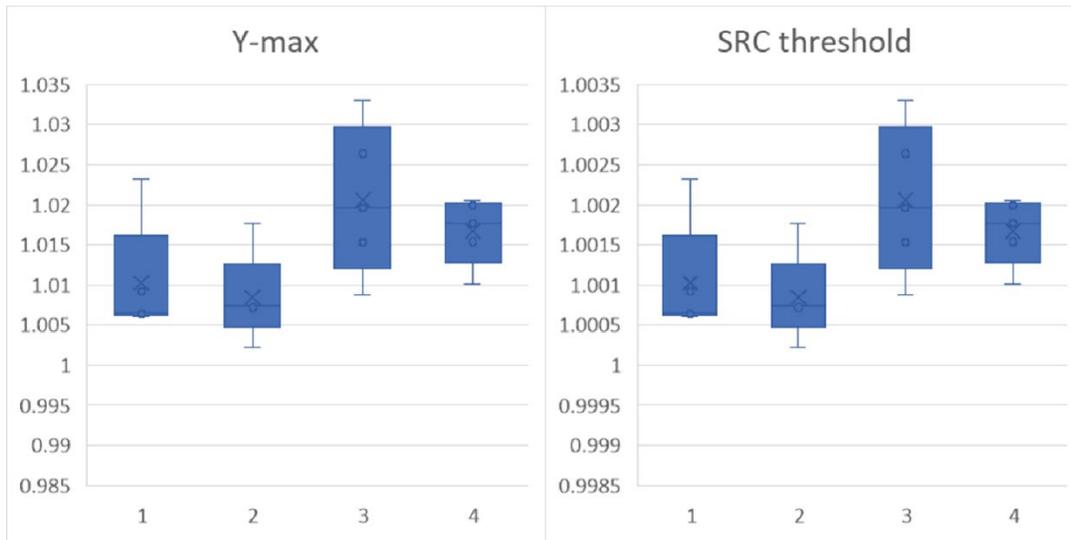

Figure 5. Significant SSFA parameters differing between individuals (exclusively between individuals #2 and #3) from Gelsight SMALL scans, including maximum relative distance (also known as *Y max*) (p = .047) and Smooth-rough crossover threshold (*SRC threshold*) (p = .047).